\documentclass[
]{revtex4-1}

\usepackage{upgreek}
\usepackage{graphicx}
\usepackage{dcolumn}
\usepackage{bm}

\begin{document}


\title[Sample title]{Broadband Magnetometry and Temperature Sensing with a Light Trapping Diamond Waveguide}



\author{Hannah Clevenson}
\email{hannahac@mit.edu}
\affiliation{Department of Electrical Engineering and Computer Science, Massachusetts Institute of Technology, Cambridge, MA 02139, USA}
\author{Matthew E. Trusheim}
\affiliation{Department of Electrical Engineering and Computer Science, Massachusetts Institute of Technology, Cambridge, MA 02139, USA}
\author{Tim Schr\"{o}der}
\affiliation{Department of Electrical Engineering and Computer Science, Massachusetts Institute of Technology, Cambridge, MA 02139, USA}
\author{Carson Teale}
\affiliation{Department of Electrical Engineering and Computer Science, Massachusetts Institute of Technology, Cambridge, MA 02139, USA}
\author{Dirk Englund}
\email{englund@mit.edu}
\affiliation{Department of Electrical Engineering and Computer Science, Massachusetts Institute of Technology, Cambridge, MA 02139, USA}
\author{Danielle Braje}
\email{braje@ll.mit.edu}
\affiliation{MIT Lincoln Laboratory, Lexington, MA 02420, USA}

\date{\today}


\maketitle

Solid-state quantum sensors are attracting wide interest because of their exceptional sensitivity at room temperature. In particular, the spin properties of individual nitrogen vacancy (NV) color centers in diamond~\cite{Balasubramanian:2009fk, stanwix2010coherence, Bar-Gill:2013fk} make it an outstanding nanoscale sensor of magnetic fields~\cite{2007.NPhys.Budker.optical_magnetometry, 2008.NPhys.Taylor, Maze2008, 2008.Nature.Wrachtrup.magnetometry_NV_etal, PhysRevLett.110.130802, rondin2013magnetometry}, electric fields~\cite{Dolde:2011uq}, and temperature~\cite{Kucsko:2013fk, toyli2013fluorescence, doi:10.1021/nl401216y, Doherty2013} under ambient conditions. Recent work on ensemble NV-based magnetometers~\cite{2010.Budker.APL, PhysRevB.85.121202, 2014.PRL.Budker}, inertial sensors \cite{PhysRevA.86.062104}, and clocks~\cite{PhysRevA.87.032118} have employed $N$ unentangled colour centers to realize a factor of up to $\sqrt{N}$ improvement in sensitivity~\cite{PhysRevB.85.121202, 1367-2630-13-4-045021}.  However, to realize fully this signal enhancement, new techniques are required to excite efficiently and to collect fluorescence from large NV ensembles. Here, we introduce a light-trapping diamond waveguide (LTDW) geometry that enables both high fluorescence collection ($\sim20\%$) and efficient pump absorption achieving an effective path length exceeding $1$~meter in a millimeter-sized device. The LTDW enables in excess of $2\%$ conversion efficiency of pump photons into optically detected magnetic resonance (ODMR) \cite{PhysRevB.74.104303} fluorescence, a \textit{three orders of magnitude} improvement over previous single-pass geometries. This dramatic enhancement of ODMR signal enables broadband measurements of magnetic field and temperature at less than $1$~Hz, a frequency range inaccessible by dynamical decoupling techniques. 
We demonstrate $ \sim 1~\mbox{nT}/\sqrt{\mbox{Hz}}$ magnetic field sensitivity for $0.1$~Hz to $10$~Hz and a thermal sensitivity of $\sim 400~\upmu\mbox{K}/\sqrt{\mbox{Hz}}$ and estimate a spin projection limit at $\sim 0.36~\mbox{fT}/\sqrt{\mbox{Hz}}$ and $\sim 139~\mbox{pK}/\sqrt{\mbox{Hz}}$, respectively.

\begin{figure} [!h]
\centering
\includegraphics[width=170mm]{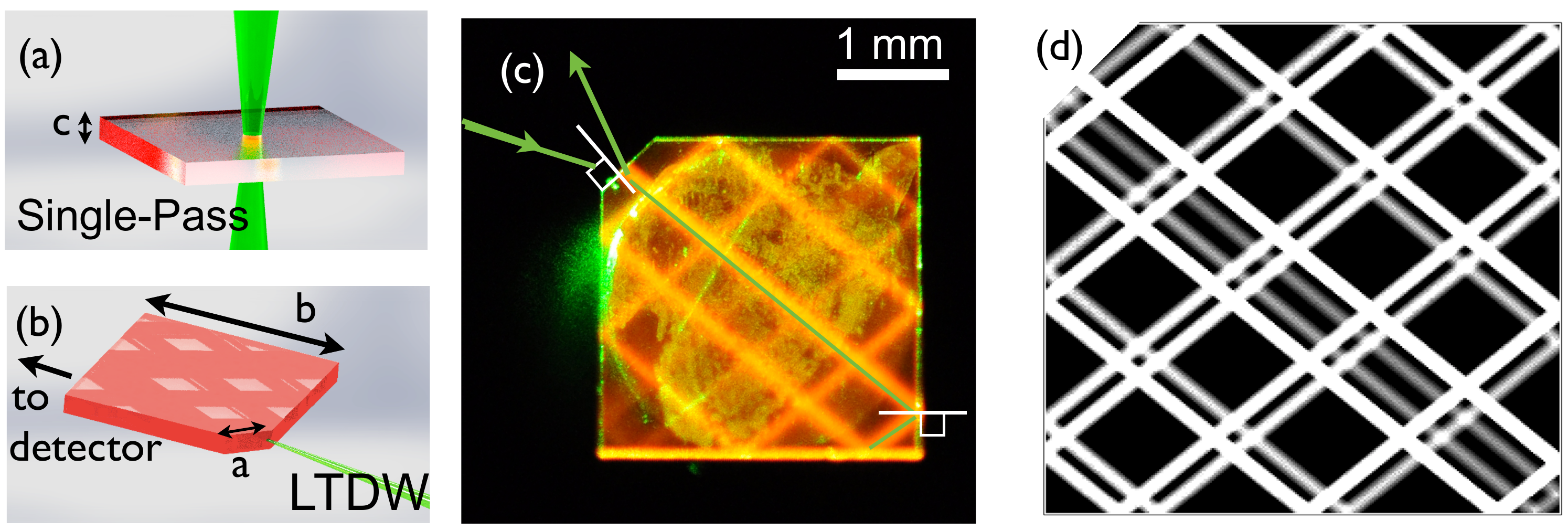}
\caption{\textbf{Light Trapping Diamond Waveguide} (a) Standard single-pass laser excitation exhibiting a short path length of interaction with NV centers as compared to the LTDW. (b) For an equivalent excitation power, the LTDW increases the path length of the excitation beam. (c) The color CCD image of the LTDW, excited by a green pump laser of 70 mW, shows bright fluorescence without any spectral filtering. (d) Simulation of the beam path in the LTDW with $a = 500~\upmu$m input facet, 3 mm side length, $100~\upmu$m diameter Gaussian profile input incident at $11.6^\circ$, and an absorption constant of $\upalpha = 0.45~\mbox{cm}^{-1}$.}
\label{levels}
\end{figure}

The NV's low absorption cross section \cite{:/content/aip/journal/jcp/129/12/10.1063/1.2987717} has thus far prevented efficient excitation and collection of ODMR signal from bulk diamond samples using the typical vertical illumination geometry shown in Fig. \ref{levels}a. For NV densities around $10^{15}~\mbox{cm}^{-3}$~\cite{2008.NPhys.Taylor, PhysRevB.87.115122}, corresponding to a nitrogen density of $10^{16}~\rm{cm}^{-3}$ assuming a nitrogen vacancy yield of 10\%, for which the NV spacing in the diamond lattice is sufficiently sparse to maintain long spin coherence times~\cite{Maze2008}, meter-long path lengths would be necessary to absorb the excitation beam (see Supplementary).  Techniques have been developed to improve fluorescence collection from single and ensemble NV systems, but the overall conversion efficiency from green pump to red fluorescence signal is still low. Efficient collection has been demonstrated from NVs excited in a single-pass geometry \cite{PhysRevB.85.121202}, but absorption is less than $1.3\%$ for a typical $300~\upmu$m thick, electron-irradiated, type IIa diamond sample.
Wide-field microscopy, implemented with charge-coupled device (CCD) cameras, can collect fluorescence from ensembles of up to $10^3$~NV centers \cite{Bar-Gill:2012fk} in the focal volume of the objective, but suffers from low collection efficiency. External Fabry-Perot cavities have been investigated to increase the optical depth of the infrared resonant transition \cite{2014.PRL.Budker}; this technique could be used for green excitation, but would introduce the additional complexity of stabilized, narrow-linewidth, green excitation lasers with bandwidth restrictions. For instance, a cavity finesse of $F\sim 1000$ would be needed to enhance the path length of a mm-scale device to one meter requiring laser stabilization to a $\sim62$~MHz linewidth.  

The LTDW consists of a square diamond slab with a small angled facet at one corner for the input-coupling of the pump beam, as shown in Fig.~\ref{levels}b. The input facet has a length $a$ at an angle $45^{\rm \circ}$ relative to the square sample sides, allowing pump light to couple into the structure while being confined by total internal reflection (TIR) ($\theta > \theta_{c}=24.6^\circ$) on the other surfaces.  Fig.~\ref{levels}c shows an optical image of a pump beam coupled at an $11.6^{\circ}$~angle into the LTDW (3 mm x 3 mm x $300~\upmu$m). The strong red fluorescence is seen on a color CCD camera even without filtering of the green pump light. Figure 1d shows the corresponding numerical simulation of a $100~\upmu$m diameter Gaussian pump laser incident on an identical LTDW structure. To illustrate the pump path without too much overlap, this figure shows a non-optimal beam path of 6~cm.
As we show below, several meters of propagation length is possible in such structures, enabling full absorption of the pump beam even for low NV density. With low NV densities and an optimized excitation pattern and geometry \cite{narimanov2005compact}, NVs throughout the LTDW can thus be excited. The fluorescence is collected at the edges of the diamond (see methods).

\begin{figure} [!h]
\centering
\includegraphics[width=170mm]{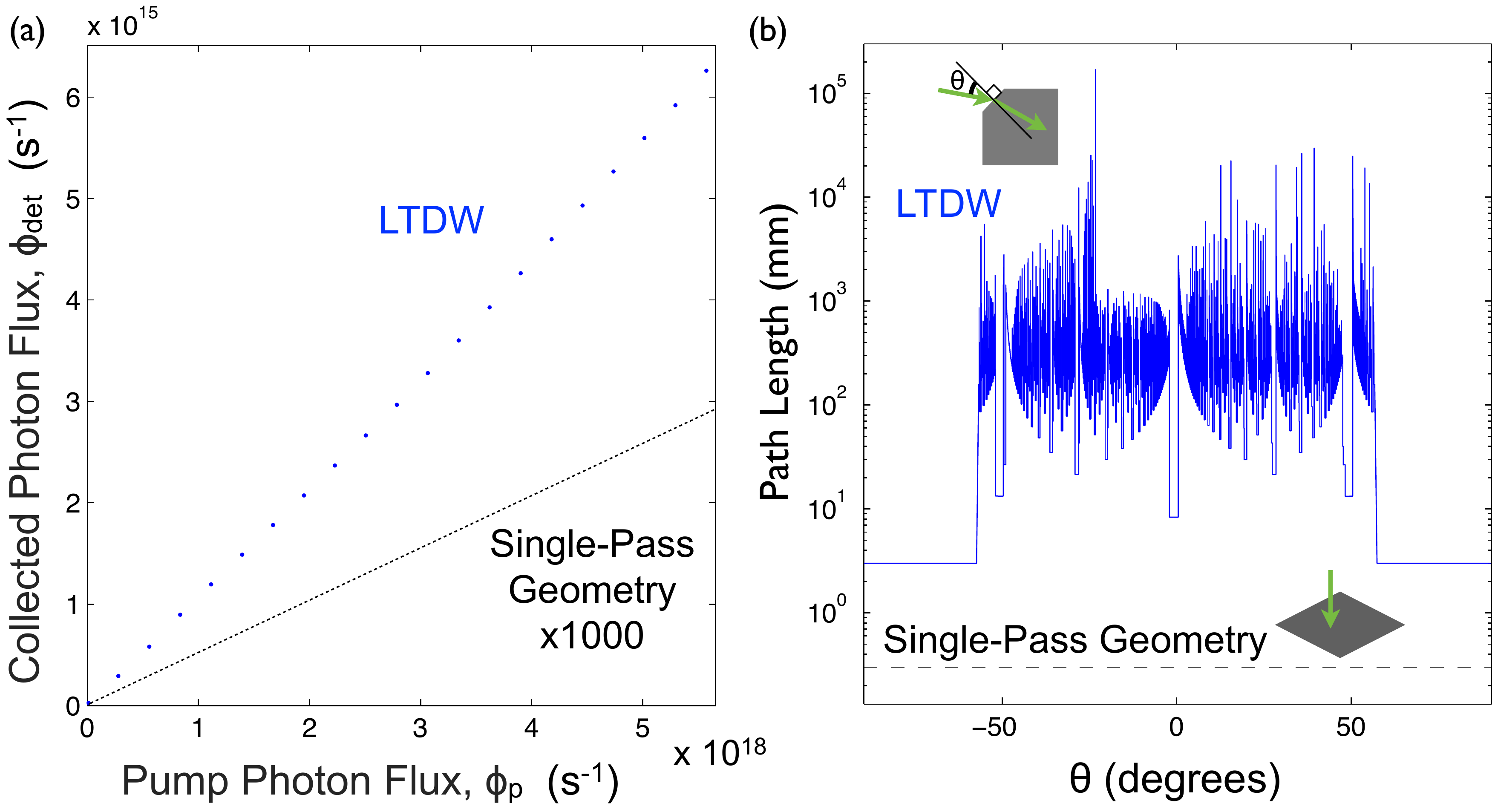}
\caption{\textbf{Optical properties of the LTDW} (a) The LTDW shows a factor of $2.3\times 10^3$ improvement compared with the expected detected photon flux from a single pass through a diamond sample of the same absorption constant and collection using a solid immersion lens and confocal setup. Error bars for the measured data (LTDW case) are contained within the data points. (b) Maximum optical path length is plotted as a function of input angle into the 3 mm x 3 mm LTDW structure with a $150~\upmu$m input facet. The excitation laser intersects the input facet $50~\upmu$m from the center, resulting in asymmetry in the plot. For comparison, the maximum optical path length achievable in a single-pass configuration is also plotted. In samples with sufficient NV density, the excitation beam would be absorbed before achieving the maximum path length allowed by the LTDW structure.}
\label{satcurve}
\end{figure}

As a central figure of merit in ensemble NV experiments, we sought to optimize the ODMR conversion efficiency, $\eta_c$, from green pump input photon flux, $\phi_{p}$, to red fluorescence flux, $\phi_{f}$, where $\eta_c=\phi_{f}/\phi_{p}$. The fluorescence flux is approximated from the absorbed photon flux, $\phi_{abs}$, by $\phi_{f}=\eta_q \phi_{abs}$, where the fluorescence quantum efficiency ($\eta_q$) is thought to be near 0.7~\cite{0953-8984-18-21-S08} for NVs in bulk diamond due to the charge state transfer to $\mbox{NV}^0$. Figure \ref{satcurve}a shows the experimentally detected photon flux $\phi_{det}$ when the LTDW input facet is pumped with a 532 nm wavelength Gaussian beam (diameter $=300 ~\upmu$m). The incident photon flux $\phi_{p}$ is obtained from the input power $ P_{p}$: $\phi_{p}=P_{p}/(\hbar \omega) \cdot t_{p}$, where the transmission, $t_{p}\sim0.83$, accounts for the  step-index reflection at the input facet and $\hbar \omega= 2.33$~eV is the green photon energy. The detection efficiency of fluorescence photons is estimated from the fraction of photons emitted into angles that are not confined by TIR ($27.2\%$ of total fluorescence), the fraction of these photons that are imaged onto the detector (20\%), and the photodetector quantum efficiency $(\sim80\%)$. This gives a detection efficiency of $\eta_{det}\sim 0.043$. The fluorescence photon flux is then $\phi_{f}=\phi_{det}/\eta_{det}$. From the slope of a linear fit to the LTDW signal in Fig. \ref{satcurve}a, we calculate the total ODMR efficiency, quantifying the conversion of pump photons to detected photons: $\eta_{ODMR} =\phi_{det}/\phi_{p}=\eta_c \cdot \eta_{det} \sim 1.032 \times 10^{-3}$, \textit{i.e}., approximately one red photon is detected for every 970 pump photons. This results in a pump-to-fluorescence conversion efficiency of $\eta_c \sim 0.024$. Compared to the calculated maximum conversion efficiency for a single-pass geometry addressing a volume of NVs with the addition of a solid immersion lens as in \cite{schroder2011ultrabright} (dashed line in Fig.~\ref{satcurve}a), the LTDW conversion efficiency $\eta_c$ is improved by over three orders of magnitude. Further improvement in $\eta_c$ could be obtained by further confining the input beam through reducing the size of the input facet. Detection efficiency could also be increased by a factor of 2.5 by directly positioning detectors on the four diamond facets \cite{PhysRevB.85.121202}, though at the expense of additional electrical noise. 

The linear relationship between $\phi_{p}$ and $\phi_{det}$ in Figure \ref{satcurve}a shows that the NV is pumped below saturation when the pump laser power is below 2 W. Assuming perfect TIR, negligible scattering loss at the diamond surfaces, and using the measured absorption constant ($\upalpha \simeq 0.45~\rm{cm^{-1}}$) and the calculated NV density ($2.3 \times 10^{16}~\rm{cm}^{-3}$, see Supplementary)of the device, more than $99\%$ of the input excitation  beam is absorbed over a path length of $15$~cm. Simulations of path length versus incident angle indicate that the LTDW structure enables path length of over a meter (see Figure \ref{satcurve}b). This increase is especially relevant for samples with low NV density and therefore lower $\upalpha$ values, which have attractive properties such as longer coherence times. 

\begin{figure} [!h]
\centering
\includegraphics[width=170mm]{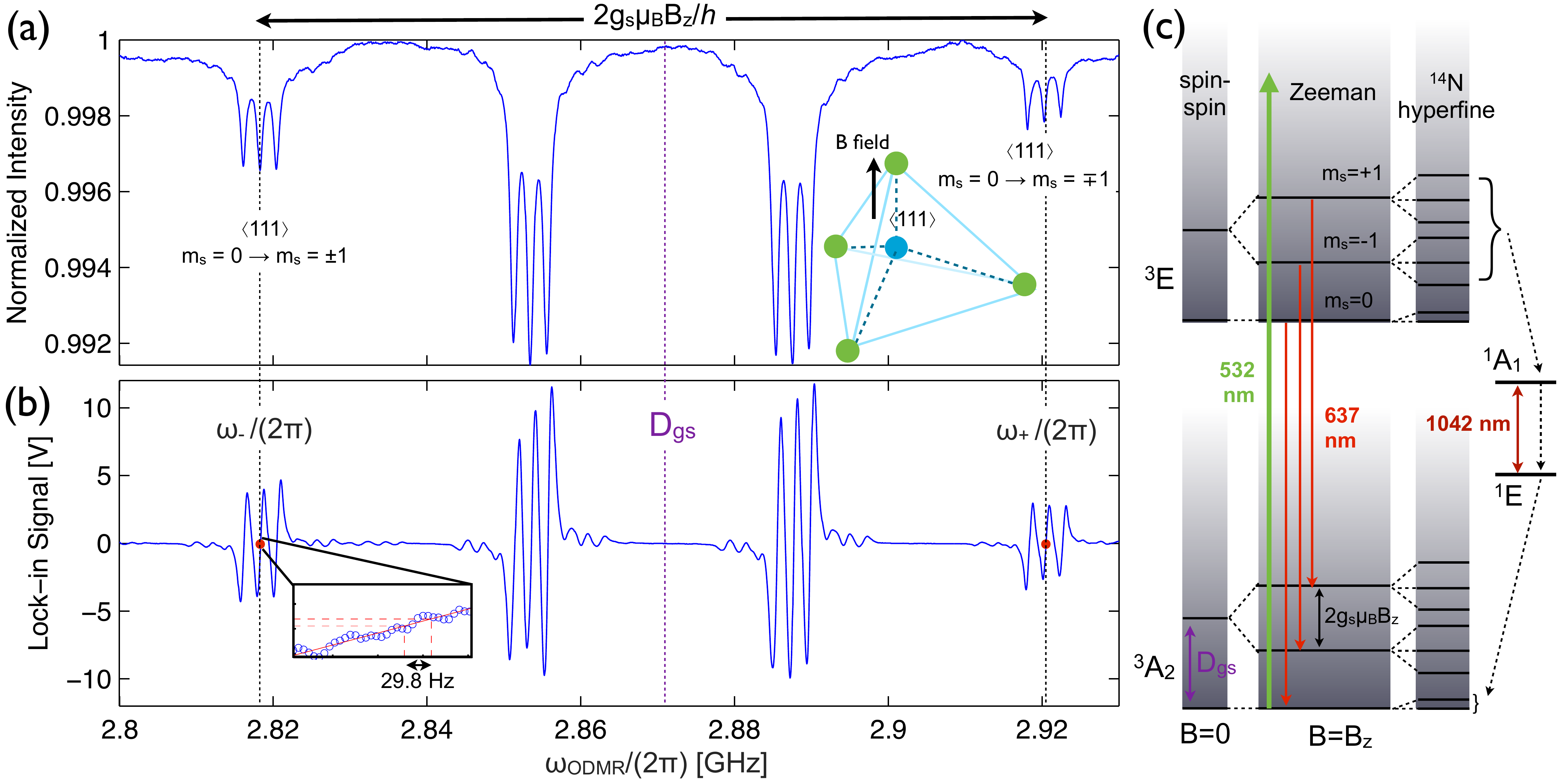}
\caption{\textbf{ODMR spectra and level structure} (a) CW electron spin resonance with external magnetic field aligned along a single $\langle 111 \rangle$ crystal axis. The inset depicts the diamond lattice, where the blue circle represents the nitrogen atom and the green circles show the four possible orientations for the vacancy in the tetrahedral crystal lattice, constituting the four sub-ensembles of NVs. (b) Lock-in output corresponding to signal in (a) with SNR of $10^5$ after $1$ second of averaging. The scale factor (V/Hz) is provided by a linear fit around each of the two red points.  Inset shows detail of the noise on this curve around the red point. Temperature and magnetic field shifts are measured independently by tracking both $m_s=\pm 1$. (c) Energy-level diagram of diamond NV center showing radiative (solid lines) and non-radiative (dotted lines) transitions. }
\label{data}
\end{figure}

The increased path length allows for high precision ODMR measurements. Figure \ref{data}a plots ODMR spectra under continuous-wave (CW) microwave field excitation using a $200~\upmu$m-diameter wire loop placed 3 mm above the LTDW. A 2.5 mT static magnetic field applied along the $\langle111\rangle$ diamond crystal axis splits the degeneracy of the NV sub-level transitions $\mbox{m}_s=0 \rightarrow \pm 1$. The $\langle 111 \rangle$ sub-ensemble of NVs, whose quantization axis is aligned with the magnetic field direction, shows the greatest splitting, as indicated in Figure \ref{data}a by the frequencies $\omega_+$ and $\omega_-$.  The equal magnetic field projection along each axis results in the transitions of the three remaining NV orientations having a degenerate frequency splitting with three times the ODMR contrast compared to the $\langle111\rangle$ sub-ensemble.  Hyperfine coupling to nearby nuclear spins splits each of the electronic sub-level transitions into Lorentzian triplets (Fig. 3c), which exhibit a full-width half-maximum of $1.2$~MHz spaced by $\sim 2.1~\rm{MHz}$.

We measure magnetic field and temperature shifts by monitoring the fluorescence of the LTDW at the $m_s=0\rightarrow -1$ and the $m_s=0 \rightarrow $+1 transitions. 
The microwave excitation frequency, $\omega_{\mathrm{ODMR}}$, is tuned alternatingly to $\omega_+$ and $\omega_-$, where the derivative of the ODMR signal with respect to frequency is greatest. We modulate $\omega_{\mathrm{ODMR}}$ at 1.5~kHz with a modulation depth of 1~MHz, and use a lock-in amplifier to monitor the NV fluorescence.  Figure \ref{data}b plots the lock-in signal for a time constant of 10 ms. The high SNR ($10^5$) enables the resolution of NV sub-level transitions to 29.8 Hz for a 1-second integration time per point (see inset of Figure 3b). 

\begin{figure} [!ht]
\centering
\includegraphics[width=170 mm]{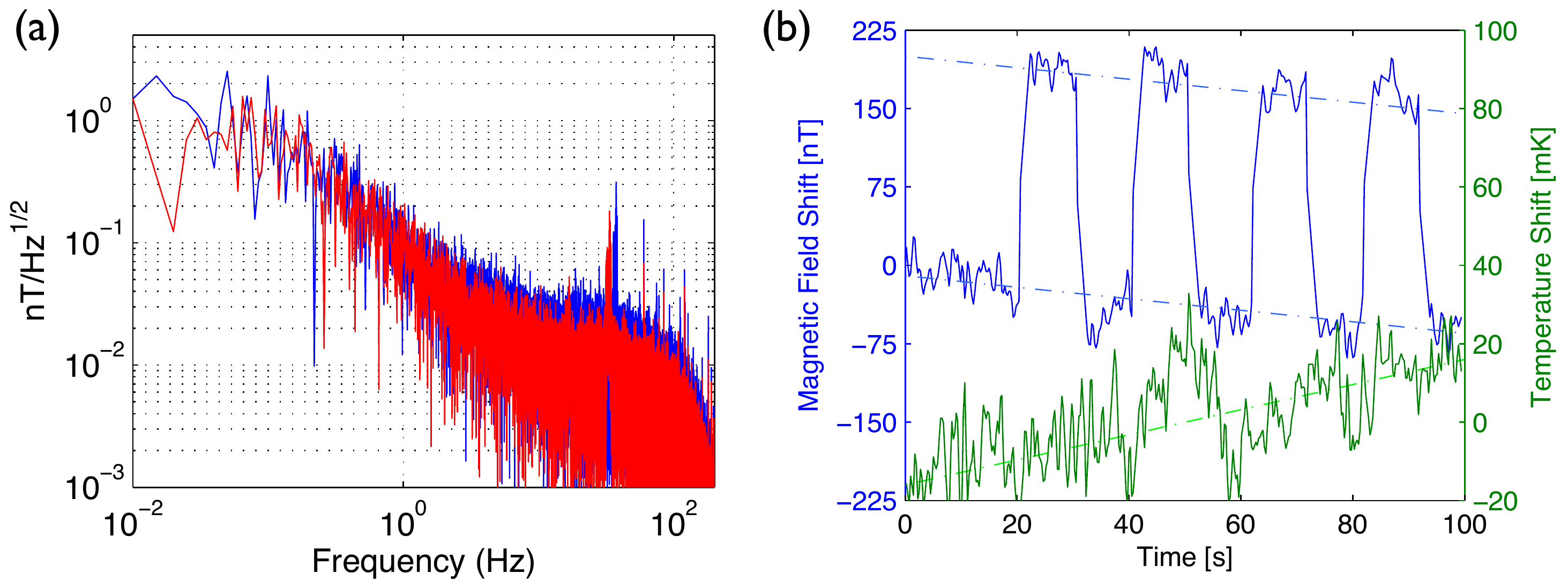}
\caption{\textbf{Sensitivity measurements} (a) Discrete Fourier transform of the lock-in amplifier noise signal for $\omega_-$ (red) and $\omega_+$ (blue) microwave resonances. (b) Separation of magnetic field and temperature effects.  Green and blue lines show the sum and difference of the error signals from lower and higher frequency resonances of the $\langle111\rangle$ orientation, decoupling the temperature and magnetic field drifts over 100~s, respectively. A $200$~nT field along the $\langle111\rangle$ quantization axis is applied as a 10~s period square wave, clearly visible in the blue trace. The green trace shows drift attributed to thermal effects.}
\label{drifts}
\end{figure} 

From the form of the ground state hyperfine Hamiltonian (see Supplementary), it follows that the $m_s = 0 \rightarrow \pm 1$ splittings are given by $ \omega_{\pm} = (D_{gs}+ \beta_T \Delta T) \pm \gamma B_z$, where $\beta_T=-74 ~\mbox{kHz}/\mbox{K}$ is the phenomenological temperature dependence of the NV center zero field splitting (ZFS) resonance at room temperature \cite{2010.PRL.Budker.temperature_dependence} and the Zeeman shift in response to an axially applied magnetic field in the $\langle111\rangle$ crystal direction is $\gamma =\upmu_B g_s/h = 28 ~\mbox{GHz}/\mbox{T}$, where $\upmu_B = e \hbar/2 m_e$ is the Bohr magneton and $g_s$ is the Land\'{e} factor. By sequentially monitoring both transitions, $\omega_{\pm}$, we obtain the mean value of $\omega_\pm$, yielding the temperature shift $\Delta T$, and the difference, $\omega_+ - \omega_-$, providing the magnetic field shift \cite{2013.Beausoleil.PRL}. 
The 29.8~Hz resolution shown in Fig. \ref{data}b thus corresponds to a $1.06$~nT minimum detectable shift in magnetic field or a $400 ~\upmu$K minimum detectable shift in temperature. Similarly, frequency analysis of the noise in the LTDW system (Fig. \ref{drifts}a) indicates that at $1$~Hz, both transitions of the $\langle111\rangle$ orientation sub-ensemble are sensitive to less than $1~ \mathrm{nT}/\sqrt{\mathrm{Hz}}$. Figure \ref{drifts}b plots independent  measurement of temperature and magnetic field over a time period of 100 seconds. Each point represents an integration time of 50 ms. We clearly resolve a 200~nT square-wave magnetic field applied in the $\langle111\rangle$ direction in addition to a $\sim75$~nT magnetic field drift (blue curve); the drift is attributed to geomagnetic fluctuations~\cite{1967geomagnetics} as these experiments were performed without magnetic shielding. We separately resolve a gradual temperature shift during this measurement (green curve), which occurs when the sample temperature is not actively stabilized. Note that for this method of temperature and magnetic field shift separation, signal error adds in quadrature for $\sim \sqrt{2}$~nT/$\sqrt{\rm{Hz}}$ resolution.

Figure 5 plots the performance of the LTDW sensor compared to recently demonstrated NV magnetometers. The system noise limit from Figure \ref{drifts}a is superimposed over the frequency range and sensitivity requirements for three key magnetometer applications: magnetocardiography, magnetoencelephography, and the monitoring of geomagnetics. The high SNR of the LTDW allows modulation between the low and high frequency resonance transitions of one NV center orientation for separating effects of temperature and magnetic field at short integration times beyond the capabilities of confocal microscopy. The unique position of the LTDW in the low-frequency limit makes it suited for a variety of magnetic field sensing applications including medical monitoring, object detection, and the study of geomagnetics which occur in the $10^{-4}$ to $10$~Hz range.

\begin{figure} [!ht]
\centering
\includegraphics[width=170mm]{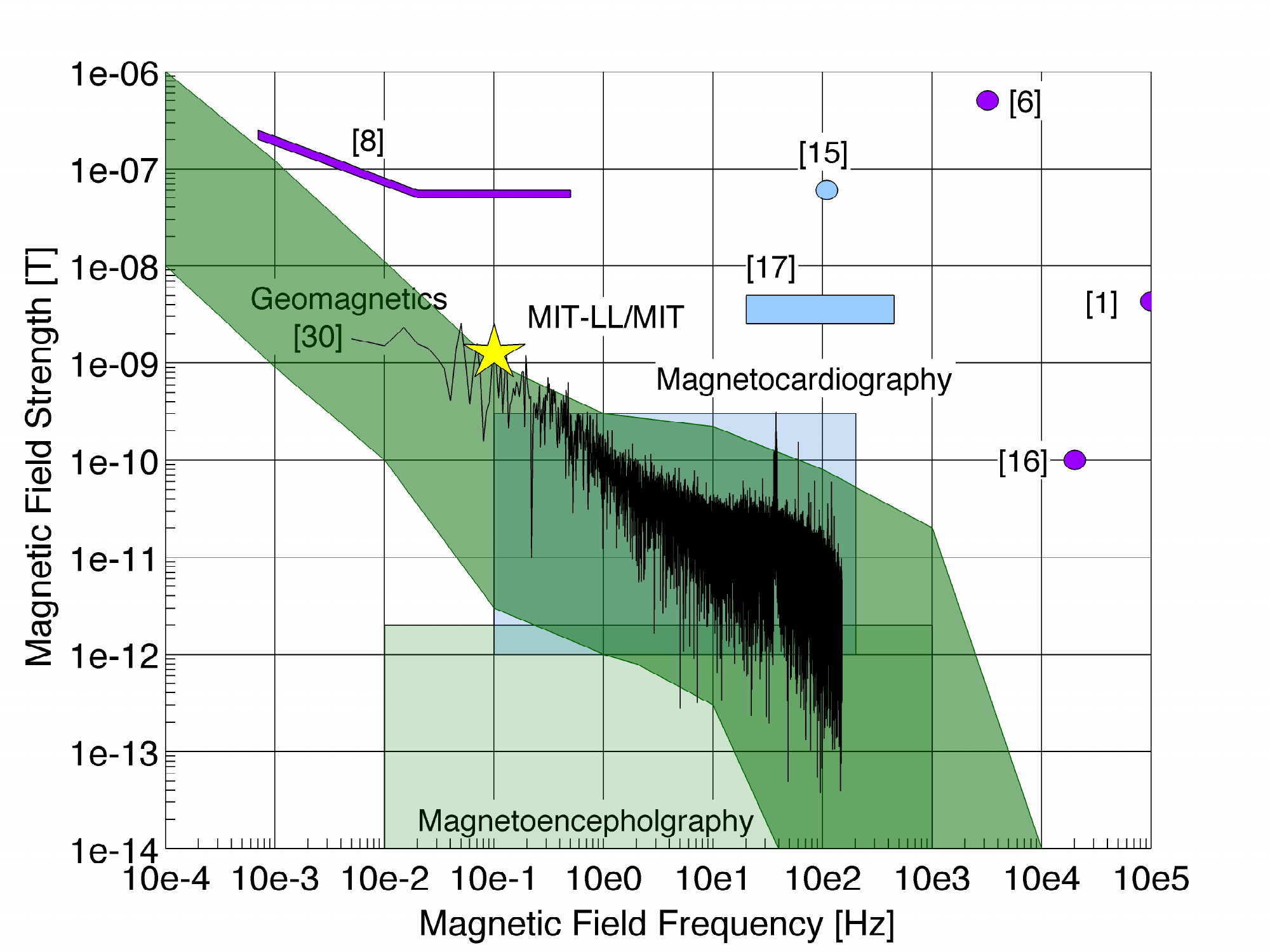}
\caption{\textbf{Comparison with other state-of-the-art diamond-based sensors} Particularly with regard to target applications such as geomagnetics~\cite{1967geomagnetics}, the LTDW operates with higher sensitivity in the most valuable frequency range. The noise spectrum overlay on this plot is representative of the system noise limits of the LTDW apparatus.}
\label{discussion}
\end{figure} 

At higher magnetic field frequencies, Ramsey-type magnetometry could be employed with dynamic decoupling pulses to extend the NV electron spin coherence time. Then, the sensitivity could approach the fundamental spin projection limit, 
\begin{equation}
\centering
S = \delta B_{\rm min} \sqrt{T} \simeq \frac{1}{\gamma}\frac{1}{\sqrt{N\tau}},
\label{limit}
\end{equation}
where $\delta B_{\rm min}$ is the minimum detectable magnetic field, $T$ is the measurement time, $\gamma=28$~GHz/T is the NV gyromagnetic ratio, $N = 10^{13}$ is the estimated total number of NV centers in our sample (see Supplementary), and $\tau\sim 1$~ms is the electron spin dephasing time measured in comparable samples and expected to be feasible for NV densities on the order of 1 ppm \cite{2008.NPhys.Taylor}. These parameters result in a spin-projection limit of $S = 0.36~\mbox{fT}/\sqrt{\mbox{Hz}}$. The sensitivity of the LTDW could be increased with improvements to the collection efficiency, which can be increased by a factor of 2.5 with the addition of prisms or mirrors directing the fluorescence from non-adjacent sides of the diamond to the detector or with multiple photodetectors. Other TIR geometries utilizing chaotic excitation paths could also be used for NV excitation and may result in more even coverage of the bulk sample \cite{narimanov2005compact}.  While the facets of the LTDW are highly polished to a surface roughness of less than 15 nm, scattering loss occurs at the edges and corners of the structure. Improvements in the polishing process would improve reflectivity and lower scattering losses. Additionally, increasing ODMR contrast via stronger microwave excitation would also result in improved sensitivity. 

In conclusion, we have introduced a dual magnetic field and temperature sensor based on a light-trapping diamond waveguide geometry that enables nearly complete green pump absorption and efficient spin-dependent fluorescence collection. With more than three orders of magnitude improvement in pump power conversion over previous single-pass schemes, the LTDW achieves high signal to noise sensing and nT/$\sqrt{\mbox{Hz}}$ sensitivity in the low-frequency regime, which is not accessible by dynamical decoupling techniques. 
With record low frequency sensitivity, the LTDW promises new applications for spin-based solid state precision sensors in a range of fields, including biomedical sensing and object detection. The low NV density allows clear resolution of the hyperfine spectra for improved sensitivity to temperature and magnetic field shifts. The light trapping technique could be extended to transmission measurements and direct IR absorption techniques~\cite{2014.PRL.Budker}. The device is not limited to sensing applications; it could be used for any application requiring a long optical path length within the sample, such as electromagnetically induced transparency experiments or optical quantum memories.
The LTDW provides a compact, portable precision sensor platform for measuring magnetic fields, temperature, pressure, rotation, or time.  Via multiplexing, any combination of these target sensing applications could be performed in a single compact device.

\section*{Methods}
\subsection*{Sample Preparation} 
We begin with a $\langle 100 \rangle$-oriented, type IIa CVD diamond produced by chemical vapor deposition (Element 6) with dimensions ($3 \times 3 \times 0.3~\rm{mm}^3$).  To increase NV ensemble densities, the diamond was electron-irradiated at $4.5$~MeV with a beam current of 20 mA over $\sim0.15~\mbox{m}^2$ and annealed for 2 hours at $850^\circ$C for $\sim 0.1$~ppm NV centers. All six surfaces are polished to a surface roughness of less than $15$~nm with a $~500~\upmu$m input facet on one corner at $45^\circ$ creating an entrance window for the laser (see Fig. \ref{levels}a).  The diamond sample is thermally anchored to a brass sample holder with indium, which may cause additional absorption through evanescent coupling of the confined radiation.

\subsection*{Experimental Setup}
A spatially filtered, $532$~nm excitation beam from a Verdi V5 is focused into the input facet of the diamond sample.  Three sets of mutually perpendicular Helmholtz coils, with diameters ranging from 8 to 12 inches, provide a uniform static or slowly varying magnetic field.  Using ODMR spectra recorded with the application of three different control fields, the rotation matrix from the cartesian frame of the coils to the NV axis is determined.  An impedance-matched loop antenna located $\sim 2.5$ mm above the sample delivers microwave excitation.  The microwave excitation is frequency modulated at $1.5$~kHz with a modulation depth of $1$~MHz.  Fluorescence emission is collected with an aspheric condenser lens (NA=0.79), followed by a $662-800$~nm bandpass filter to minimize NV$^0$ contribution to the fluorescence. The system collection efficiency is $\sim 20\%$ of the total fluorescence signal.  A second condenser lens images the fluorescence onto an amplified 9.8 mm-diameter Si transimpedence amplified photodetector.  The signal is demodulated using an SRS-850 lock-in amplifier.


\begin{thebibliography}{30}%
\makeatletter
\providecommand \@ifxundefined [1]{%
 \@ifx{#1\undefined}
}%
\providecommand \@ifnum [1]{%
 \ifnum #1\expandafter \@firstoftwo
 \else \expandafter \@secondoftwo
 \fi
}%
\providecommand \@ifx [1]{%
 \ifx #1\expandafter \@firstoftwo
 \else \expandafter \@secondoftwo
 \fi
}%
\providecommand \natexlab [1]{#1}%
\providecommand \enquote  [1]{``#1''}%
\providecommand \bibnamefont  [1]{#1}%
\providecommand \bibfnamefont [1]{#1}%
\providecommand \citenamefont [1]{#1}%
\providecommand \href@noop [0]{\@secondoftwo}%
\providecommand \href [0]{\begingroup \@sanitize@url \@href}%
\providecommand \@href[1]{\@@startlink{#1}\@@href}%
\providecommand \@@href[1]{\endgroup#1\@@endlink}%
\providecommand \@sanitize@url [0]{\catcode `\\12\catcode `\$12\catcode
  `\&12\catcode `\#12\catcode `\^12\catcode `\_12\catcode `\%12\relax}%
\providecommand \@@startlink[1]{}%
\providecommand \@@endlink[0]{}%
\providecommand \url  [0]{\begingroup\@sanitize@url \@url }%
\providecommand \@url [1]{\endgroup\@href {#1}{\urlprefix }}%
\providecommand \urlprefix  [0]{URL }%
\providecommand \Eprint [0]{\href }%
\providecommand \doibase [0]{http://dx.doi.org/}%
\providecommand \selectlanguage [0]{\@gobble}%
\providecommand \bibinfo  [0]{\@secondoftwo}%
\providecommand \bibfield  [0]{\@secondoftwo}%
\providecommand \translation [1]{[#1]}%
\providecommand \BibitemOpen [0]{}%
\providecommand \bibitemStop [0]{}%
\providecommand \bibitemNoStop [0]{.\EOS\space}%
\providecommand \EOS [0]{\spacefactor3000\relax}%
\providecommand \BibitemShut  [1]{\csname bibitem#1\endcsname}%
\let\auto@bib@innerbib\@empty
\bibitem [{\citenamefont {Balasubramanian}\ \emph {et~al.}(2009)\citenamefont
  {Balasubramanian}, \citenamefont {Neumann}, \citenamefont {Twitchen},
  \citenamefont {Markham}, \citenamefont {Kolesov}, \citenamefont {Mizuochi},
  \citenamefont {Isoya}, \citenamefont {Achard}, \citenamefont {Beck},
  \citenamefont {Tissler}, \citenamefont {Jacques}, \citenamefont {Hemmer},
  \citenamefont {Jelezko},\ and\ \citenamefont
  {Wrachtrup}}]{Balasubramanian:2009fk}%
  \BibitemOpen
  \bibfield  {author} {\bibinfo {author} {\bibfnamefont {G.}~\bibnamefont
  {Balasubramanian}}, \bibinfo {author} {\bibfnamefont {P.}~\bibnamefont
  {Neumann}}, \bibinfo {author} {\bibfnamefont {D.}~\bibnamefont {Twitchen}},
  \bibinfo {author} {\bibfnamefont {M.}~\bibnamefont {Markham}}, \bibinfo
  {author} {\bibfnamefont {R.}~\bibnamefont {Kolesov}}, \bibinfo {author}
  {\bibfnamefont {N.}~\bibnamefont {Mizuochi}}, \bibinfo {author}
  {\bibfnamefont {J.}~\bibnamefont {Isoya}}, \bibinfo {author} {\bibfnamefont
  {J.}~\bibnamefont {Achard}}, \bibinfo {author} {\bibfnamefont
  {J.}~\bibnamefont {Beck}}, \bibinfo {author} {\bibfnamefont {J.}~\bibnamefont
  {Tissler}}, \bibinfo {author} {\bibfnamefont {V.}~\bibnamefont {Jacques}},
  \bibinfo {author} {\bibfnamefont {P.~R.}\ \bibnamefont {Hemmer}}, \bibinfo
  {author} {\bibfnamefont {F.}~\bibnamefont {Jelezko}}, \ and\ \bibinfo
  {author} {\bibfnamefont {J.}~\bibnamefont {Wrachtrup}},\ }\href
  {http://dx.doi.org/10.1038/nmat2420} {\bibfield  {journal} {\bibinfo
  {journal} {Nat Mater}\ }\textbf {\bibinfo {volume} {8}},\ \bibinfo {pages}
  {383} (\bibinfo {year} {2009})}\BibitemShut {NoStop}%
\bibitem [{\citenamefont {Stanwix}\ \emph {et~al.}(2010)\citenamefont
  {Stanwix}, \citenamefont {Pham}, \citenamefont {Maze}, \citenamefont
  {Le~Sage}, \citenamefont {Yeung}, \citenamefont {Cappellaro}, \citenamefont
  {Hemmer}, \citenamefont {Yacoby}, \citenamefont {Lukin},\ and\ \citenamefont
  {Walsworth}}]{stanwix2010coherence}%
  \BibitemOpen
  \bibfield  {author} {\bibinfo {author} {\bibfnamefont {P.~L.}\ \bibnamefont
  {Stanwix}}, \bibinfo {author} {\bibfnamefont {L.~M.}\ \bibnamefont {Pham}},
  \bibinfo {author} {\bibfnamefont {J.~R.}\ \bibnamefont {Maze}}, \bibinfo
  {author} {\bibfnamefont {D.}~\bibnamefont {Le~Sage}}, \bibinfo {author}
  {\bibfnamefont {T.~K.}\ \bibnamefont {Yeung}}, \bibinfo {author}
  {\bibfnamefont {P.}~\bibnamefont {Cappellaro}}, \bibinfo {author}
  {\bibfnamefont {P.~R.}\ \bibnamefont {Hemmer}}, \bibinfo {author}
  {\bibfnamefont {A.}~\bibnamefont {Yacoby}}, \bibinfo {author} {\bibfnamefont
  {M.~D.}\ \bibnamefont {Lukin}}, \ and\ \bibinfo {author} {\bibfnamefont
  {R.~L.}\ \bibnamefont {Walsworth}},\ }\href@noop {} {\bibfield  {journal}
  {\bibinfo  {journal} {Physical Review B}\ }\textbf {\bibinfo {volume} {82}},\
  \bibinfo {pages} {201201} (\bibinfo {year} {2010})}\BibitemShut {NoStop}%
\bibitem [{\citenamefont {Bar-Gill}\ \emph {et~al.}(2013)\citenamefont
  {Bar-Gill}, \citenamefont {Pham}, \citenamefont {Jarmola}, \citenamefont
  {Budker},\ and\ \citenamefont {Walsworth}}]{Bar-Gill:2013fk}%
  \BibitemOpen
  \bibfield  {author} {\bibinfo {author} {\bibfnamefont {N.}~\bibnamefont
  {Bar-Gill}}, \bibinfo {author} {\bibfnamefont {L.~M.}\ \bibnamefont {Pham}},
  \bibinfo {author} {\bibfnamefont {A.}~\bibnamefont {Jarmola}}, \bibinfo
  {author} {\bibfnamefont {D.}~\bibnamefont {Budker}}, \ and\ \bibinfo {author}
  {\bibfnamefont {R.~L.}\ \bibnamefont {Walsworth}},\ }\href
  {http://dx.doi.org/10.1038/ncomms2771} {\bibfield  {journal} {\bibinfo
  {journal} {Nat Commun}\ }\textbf {\bibinfo {volume} {4}} (\bibinfo {year}
  {2013})}\BibitemShut {NoStop}%
\bibitem [{\citenamefont {Budker}\ and\ \citenamefont
  {Romalis}(2007)}]{2007.NPhys.Budker.optical_magnetometry}%
  \BibitemOpen
  \bibfield  {author} {\bibinfo {author} {\bibfnamefont {D.}~\bibnamefont
  {Budker}}\ and\ \bibinfo {author} {\bibfnamefont {M.}~\bibnamefont
  {Romalis}},\ }\href {\doibase 10.1038/nphys566} {\bibfield  {journal}
  {\bibinfo  {journal} {Nature Physics}\ }\textbf {\bibinfo {volume} {3}},\
  \bibinfo {pages} {227} (\bibinfo {year} {2007})}\BibitemShut {NoStop}%
\bibitem [{\citenamefont {Taylor}\ \emph {et~al.}(2008)\citenamefont {Taylor},
  \citenamefont {Cappellaro}, \citenamefont {Childress}, \citenamefont {Jiang},
  \citenamefont {Budker}, \citenamefont {Hemmer}, \citenamefont {Yacoby},
  \citenamefont {Walsworth},\ and\ \citenamefont {Lukin}}]{2008.NPhys.Taylor}%
  \BibitemOpen
  \bibfield  {author} {\bibinfo {author} {\bibfnamefont {J.~M.}\ \bibnamefont
  {Taylor}}, \bibinfo {author} {\bibfnamefont {P.}~\bibnamefont {Cappellaro}},
  \bibinfo {author} {\bibfnamefont {L.}~\bibnamefont {Childress}}, \bibinfo
  {author} {\bibfnamefont {L.}~\bibnamefont {Jiang}}, \bibinfo {author}
  {\bibfnamefont {D.}~\bibnamefont {Budker}}, \bibinfo {author} {\bibfnamefont
  {P.~R.}\ \bibnamefont {Hemmer}}, \bibinfo {author} {\bibfnamefont
  {A.}~\bibnamefont {Yacoby}}, \bibinfo {author} {\bibfnamefont
  {R.}~\bibnamefont {Walsworth}}, \ and\ \bibinfo {author} {\bibfnamefont
  {M.~D.}\ \bibnamefont {Lukin}},\ }\href {\doibase 10.1038/nphys1075}
  {\bibfield  {journal} {\bibinfo  {journal} {Nature Physics}\ }\textbf
  {\bibinfo {volume} {4}},\ \bibinfo {pages} {810} (\bibinfo {year}
  {2008})}\BibitemShut {NoStop}%
\bibitem [{\citenamefont {Maze}\ \emph {et~al.}(2008)\citenamefont {Maze},
  \citenamefont {Stanwix}, \citenamefont {Hodges}, \citenamefont {Hong},
  \citenamefont {Taylor}, \citenamefont {Cappellaro}, \citenamefont {Jiang},
  \citenamefont {Dutt}, \citenamefont {Togan}, \citenamefont {Zibrov},
  \citenamefont {Yacoby}, \citenamefont {Walsworth},\ and\ \citenamefont
  {Lukin}}]{Maze2008}%
  \BibitemOpen
  \bibfield  {author} {\bibinfo {author} {\bibfnamefont {J.~R.}\ \bibnamefont
  {Maze}}, \bibinfo {author} {\bibfnamefont {P.~L.}\ \bibnamefont {Stanwix}},
  \bibinfo {author} {\bibfnamefont {J.~S.}\ \bibnamefont {Hodges}}, \bibinfo
  {author} {\bibfnamefont {S.}~\bibnamefont {Hong}}, \bibinfo {author}
  {\bibfnamefont {J.~M.}\ \bibnamefont {Taylor}}, \bibinfo {author}
  {\bibfnamefont {P.}~\bibnamefont {Cappellaro}}, \bibinfo {author}
  {\bibfnamefont {L.}~\bibnamefont {Jiang}}, \bibinfo {author} {\bibfnamefont
  {M.~V.~G.}\ \bibnamefont {Dutt}}, \bibinfo {author} {\bibfnamefont
  {E.}~\bibnamefont {Togan}}, \bibinfo {author} {\bibfnamefont {a.~S.}\
  \bibnamefont {Zibrov}}, \bibinfo {author} {\bibfnamefont {a.}~\bibnamefont
  {Yacoby}}, \bibinfo {author} {\bibfnamefont {R.~L.}\ \bibnamefont
  {Walsworth}}, \ and\ \bibinfo {author} {\bibfnamefont {M.~D.}\ \bibnamefont
  {Lukin}},\ }\href {\doibase 10.1038/nature07279} {\bibfield  {journal}
  {\bibinfo  {journal} {Nature}\ }\textbf {\bibinfo {volume} {455}},\ \bibinfo
  {pages} {644} (\bibinfo {year} {2008})}\BibitemShut {NoStop}%
\bibitem [{\citenamefont {Balasubramanian}\ \emph {et~al.}(2008)\citenamefont
  {Balasubramanian} \emph
  {et~al.}}]{2008.Nature.Wrachtrup.magnetometry_NV_etal}%
  \BibitemOpen
  \bibfield  {author} {\bibinfo {author} {\bibfnamefont {G.}~\bibnamefont
  {Balasubramanian}} \emph {et~al.},\ }\href {\doibase 10.1038/nature07278}
  {\bibfield  {journal} {\bibinfo  {journal} {Nature}\ }\textbf {\bibinfo
  {volume} {455}},\ \bibinfo {pages} {648} (\bibinfo {year}
  {2008})}\BibitemShut {NoStop}%
\bibitem [{\citenamefont {Fang}\ \emph
  {et~al.}(2013{\natexlab{a}})\citenamefont {Fang}, \citenamefont {Acosta},
  \citenamefont {Santori}, \citenamefont {Huang}, \citenamefont {Itoh},
  \citenamefont {Watanabe}, \citenamefont {Shikata},\ and\ \citenamefont
  {Beausoleil}}]{PhysRevLett.110.130802}%
  \BibitemOpen
  \bibfield  {author} {\bibinfo {author} {\bibfnamefont {K.}~\bibnamefont
  {Fang}}, \bibinfo {author} {\bibfnamefont {V.~M.}\ \bibnamefont {Acosta}},
  \bibinfo {author} {\bibfnamefont {C.}~\bibnamefont {Santori}}, \bibinfo
  {author} {\bibfnamefont {Z.}~\bibnamefont {Huang}}, \bibinfo {author}
  {\bibfnamefont {K.~M.}\ \bibnamefont {Itoh}}, \bibinfo {author}
  {\bibfnamefont {H.}~\bibnamefont {Watanabe}}, \bibinfo {author}
  {\bibfnamefont {S.}~\bibnamefont {Shikata}}, \ and\ \bibinfo {author}
  {\bibfnamefont {R.~G.}\ \bibnamefont {Beausoleil}},\ }\href {\doibase
  10.1103/PhysRevLett.110.130802} {\bibfield  {journal} {\bibinfo  {journal}
  {Phys. Rev. Lett.}\ }\textbf {\bibinfo {volume} {110}},\ \bibinfo {pages}
  {130802} (\bibinfo {year} {2013}{\natexlab{a}})}\BibitemShut {NoStop}%
\bibitem [{\citenamefont {Rondin}\ \emph {et~al.}(2013)\citenamefont {Rondin},
  \citenamefont {Tetienne}, \citenamefont {Hingant}, \citenamefont {Roch},
  \citenamefont {Malentinsky},\ and\ \citenamefont
  {Jacques}}]{rondin2013magnetometry}%
  \BibitemOpen
  \bibfield  {author} {\bibinfo {author} {\bibfnamefont {L.}~\bibnamefont
  {Rondin}}, \bibinfo {author} {\bibfnamefont {J.-P.}\ \bibnamefont
  {Tetienne}}, \bibinfo {author} {\bibfnamefont {T.}~\bibnamefont {Hingant}},
  \bibinfo {author} {\bibfnamefont {J.-F.}\ \bibnamefont {Roch}}, \bibinfo
  {author} {\bibfnamefont {P.}~\bibnamefont {Malentinsky}}, \ and\ \bibinfo
  {author} {\bibfnamefont {V.}~\bibnamefont {Jacques}},\ }\href@noop {}
  {\bibfield  {journal} {\bibinfo  {journal} {arXiv preprint arXiv:1311.5214}\
  } (\bibinfo {year} {2013})}\BibitemShut {NoStop}%
\bibitem [{\citenamefont {Dolde}\ \emph {et~al.}(2011)\citenamefont {Dolde},
  \citenamefont {Fedder}, \citenamefont {Doherty}, \citenamefont {Nobauer},
  \citenamefont {Rempp}, \citenamefont {Balasubramanian}, \citenamefont {Wolf},
  \citenamefont {Reinhard}, \citenamefont {Hollenberg}, \citenamefont
  {Jelezko},\ and\ \citenamefont {Wrachtrup}}]{Dolde:2011uq}%
  \BibitemOpen
  \bibfield  {author} {\bibinfo {author} {\bibfnamefont {F.}~\bibnamefont
  {Dolde}}, \bibinfo {author} {\bibfnamefont {H.}~\bibnamefont {Fedder}},
  \bibinfo {author} {\bibfnamefont {M.~W.}\ \bibnamefont {Doherty}}, \bibinfo
  {author} {\bibfnamefont {T.}~\bibnamefont {Nobauer}}, \bibinfo {author}
  {\bibfnamefont {F.}~\bibnamefont {Rempp}}, \bibinfo {author} {\bibfnamefont
  {G.}~\bibnamefont {Balasubramanian}}, \bibinfo {author} {\bibfnamefont
  {T.}~\bibnamefont {Wolf}}, \bibinfo {author} {\bibfnamefont {F.}~\bibnamefont
  {Reinhard}}, \bibinfo {author} {\bibfnamefont {L.~C.~L.}\ \bibnamefont
  {Hollenberg}}, \bibinfo {author} {\bibfnamefont {F.}~\bibnamefont {Jelezko}},
  \ and\ \bibinfo {author} {\bibfnamefont {J.}~\bibnamefont {Wrachtrup}},\
  }\href {http://dx.doi.org/10.1038/nphys1969} {\bibfield  {journal} {\bibinfo
  {journal} {Nat Phys}\ }\textbf {\bibinfo {volume} {7}},\ \bibinfo {pages}
  {459} (\bibinfo {year} {2011})}\BibitemShut {NoStop}%
\bibitem [{\citenamefont {Kucsko}\ \emph {et~al.}(2013)\citenamefont {Kucsko},
  \citenamefont {Maurer}, \citenamefont {Yao}, \citenamefont {Kubo},
  \citenamefont {Noh}, \citenamefont {Lo}, \citenamefont {Park},\ and\
  \citenamefont {Lukin}}]{Kucsko:2013fk}%
  \BibitemOpen
  \bibfield  {author} {\bibinfo {author} {\bibfnamefont {G.}~\bibnamefont
  {Kucsko}}, \bibinfo {author} {\bibfnamefont {P.~C.}\ \bibnamefont {Maurer}},
  \bibinfo {author} {\bibfnamefont {N.~Y.}\ \bibnamefont {Yao}}, \bibinfo
  {author} {\bibfnamefont {M.}~\bibnamefont {Kubo}}, \bibinfo {author}
  {\bibfnamefont {H.~J.}\ \bibnamefont {Noh}}, \bibinfo {author} {\bibfnamefont
  {P.~K.}\ \bibnamefont {Lo}}, \bibinfo {author} {\bibfnamefont
  {H.}~\bibnamefont {Park}}, \ and\ \bibinfo {author} {\bibfnamefont {M.~D.}\
  \bibnamefont {Lukin}},\ }\href {http://dx.doi.org/10.1038/nature12373}
  {\bibfield  {journal} {\bibinfo  {journal} {Nature}\ }\textbf {\bibinfo
  {volume} {500}},\ \bibinfo {pages} {54} (\bibinfo {year} {2013})}\BibitemShut
  {NoStop}%
\bibitem [{\citenamefont {Toyli}\ \emph {et~al.}(2013)\citenamefont {Toyli},
  \citenamefont {Charles}, \citenamefont {Christle}, \citenamefont
  {Dobrovitski},\ and\ \citenamefont {Awschalom}}]{toyli2013fluorescence}%
  \BibitemOpen
  \bibfield  {author} {\bibinfo {author} {\bibfnamefont {D.~M.}\ \bibnamefont
  {Toyli}}, \bibinfo {author} {\bibfnamefont {F.}~\bibnamefont {Charles}},
  \bibinfo {author} {\bibfnamefont {D.~J.}\ \bibnamefont {Christle}}, \bibinfo
  {author} {\bibfnamefont {V.~V.}\ \bibnamefont {Dobrovitski}}, \ and\ \bibinfo
  {author} {\bibfnamefont {D.~D.}\ \bibnamefont {Awschalom}},\ }\href@noop {}
  {\bibfield  {journal} {\bibinfo  {journal} {Proceedings of the National
  Academy of Sciences}\ }\textbf {\bibinfo {volume} {110}},\ \bibinfo {pages}
  {8417} (\bibinfo {year} {2013})}\BibitemShut {NoStop}%
\bibitem [{\citenamefont {Neumann}\ \emph {et~al.}(2013)\citenamefont
  {Neumann}, \citenamefont {Jakobi}, \citenamefont {Dolde}, \citenamefont
  {Burk}, \citenamefont {Reuter}, \citenamefont {Waldherr}, \citenamefont
  {Honert}, \citenamefont {Wolf}, \citenamefont {Brunner}, \citenamefont
  {Shim}, \citenamefont {Suter}, \citenamefont {Sumiya}, \citenamefont
  {Isoya},\ and\ \citenamefont {Wrachtrup}}]{doi:10.1021/nl401216y}%
  \BibitemOpen
  \bibfield  {author} {\bibinfo {author} {\bibfnamefont {P.}~\bibnamefont
  {Neumann}}, \bibinfo {author} {\bibfnamefont {I.}~\bibnamefont {Jakobi}},
  \bibinfo {author} {\bibfnamefont {F.}~\bibnamefont {Dolde}}, \bibinfo
  {author} {\bibfnamefont {C.}~\bibnamefont {Burk}}, \bibinfo {author}
  {\bibfnamefont {R.}~\bibnamefont {Reuter}}, \bibinfo {author} {\bibfnamefont
  {G.}~\bibnamefont {Waldherr}}, \bibinfo {author} {\bibfnamefont
  {J.}~\bibnamefont {Honert}}, \bibinfo {author} {\bibfnamefont
  {T.}~\bibnamefont {Wolf}}, \bibinfo {author} {\bibfnamefont {A.}~\bibnamefont
  {Brunner}}, \bibinfo {author} {\bibfnamefont {J.~H.}\ \bibnamefont {Shim}},
  \bibinfo {author} {\bibfnamefont {D.}~\bibnamefont {Suter}}, \bibinfo
  {author} {\bibfnamefont {H.}~\bibnamefont {Sumiya}}, \bibinfo {author}
  {\bibfnamefont {J.}~\bibnamefont {Isoya}}, \ and\ \bibinfo {author}
  {\bibfnamefont {J.}~\bibnamefont {Wrachtrup}},\ }\href {\doibase
  10.1021/nl401216y} {\bibfield  {journal} {\bibinfo  {journal} {Nano Letters}\
  }\textbf {\bibinfo {volume} {13}},\ \bibinfo {pages} {2738} (\bibinfo {year}
  {2013})} \BibitemShut {NoStop}%
\bibitem [{\citenamefont {Doherty}\ \emph {et~al.}(2013)\citenamefont
  {Doherty}, \citenamefont {Acosta}, \citenamefont {Jarmola}, \citenamefont
  {Barson}, \citenamefont {Manson}, \citenamefont {Budker},\ and\ \citenamefont
  {Hollenberg}}]{Doherty2013}%
  \BibitemOpen
  \bibfield  {author} {\bibinfo {author} {\bibfnamefont {M.~W.}\ \bibnamefont
  {Doherty}}, \bibinfo {author} {\bibfnamefont {V.~M.}\ \bibnamefont {Acosta}},
  \bibinfo {author} {\bibfnamefont {A.}~\bibnamefont {Jarmola}}, \bibinfo
  {author} {\bibfnamefont {M.~S.~J.}\ \bibnamefont {Barson}}, \bibinfo {author}
  {\bibfnamefont {N.~B.}\ \bibnamefont {Manson}}, \bibinfo {author}
  {\bibfnamefont {D.}~\bibnamefont {Budker}}, \ and\ \bibinfo {author}
  {\bibfnamefont {L.~C.~L.}\ \bibnamefont {Hollenberg}},\ }\href
  {http://arxiv.org/abs/1310.7303} {\bibfield  {journal} {\bibinfo  {journal}
  {arXiv preprint arXiv:1310.7303}\ } (\bibinfo {year} {2013})}\BibitemShut
  {NoStop}%
\bibitem [{\citenamefont {Acosta}\ \emph
  {et~al.}(2010{\natexlab{a}})\citenamefont {Acosta}, \citenamefont {Bauch},
  \citenamefont {Jarmola}, \citenamefont {Zipp}, \citenamefont {Ledbetter},\
  and\ \citenamefont {Budker}}]{2010.Budker.APL}%
  \BibitemOpen
  \bibfield  {author} {\bibinfo {author} {\bibfnamefont {V.}~\bibnamefont
  {Acosta}}, \bibinfo {author} {\bibfnamefont {E.}~\bibnamefont {Bauch}},
  \bibinfo {author} {\bibfnamefont {A.}~\bibnamefont {Jarmola}}, \bibinfo
  {author} {\bibfnamefont {L.~J.}\ \bibnamefont {Zipp}}, \bibinfo {author}
  {\bibfnamefont {M.}~\bibnamefont {Ledbetter}}, \ and\ \bibinfo {author}
  {\bibfnamefont {D.}~\bibnamefont {Budker}},\ }\href {\doibase
  10.1063/1.3507884} {\bibfield  {journal} {\bibinfo  {journal} {Applied
  Physics Letters}\ }\textbf {\bibinfo {volume} {97}},\ \bibinfo {pages}
  {174104} (\bibinfo {year} {2010}{\natexlab{a}})}\BibitemShut {NoStop}%
\bibitem [{\citenamefont {Le~Sage}\ \emph {et~al.}(2012)\citenamefont
  {Le~Sage}, \citenamefont {Pham}, \citenamefont {Bar-Gill}, \citenamefont
  {Belthangady}, \citenamefont {Lukin}, \citenamefont {Yacoby},\ and\
  \citenamefont {Walsworth}}]{PhysRevB.85.121202}%
  \BibitemOpen
  \bibfield  {author} {\bibinfo {author} {\bibfnamefont {D.}~\bibnamefont
  {Le~Sage}}, \bibinfo {author} {\bibfnamefont {L.~M.}\ \bibnamefont {Pham}},
  \bibinfo {author} {\bibfnamefont {N.}~\bibnamefont {Bar-Gill}}, \bibinfo
  {author} {\bibfnamefont {C.}~\bibnamefont {Belthangady}}, \bibinfo {author}
  {\bibfnamefont {M.~D.}\ \bibnamefont {Lukin}}, \bibinfo {author}
  {\bibfnamefont {A.}~\bibnamefont {Yacoby}}, \ and\ \bibinfo {author}
  {\bibfnamefont {R.~L.}\ \bibnamefont {Walsworth}},\ }\href {\doibase
  10.1103/PhysRevB.85.121202} {\bibfield  {journal} {\bibinfo  {journal} {Phys.
  Rev. B}\ }\textbf {\bibinfo {volume} {85}},\ \bibinfo {pages} {121202}
  (\bibinfo {year} {2012})}\BibitemShut {NoStop}%
\bibitem [{\citenamefont {Jensen}\ \emph {et~al.}(2014)\citenamefont {Jensen},
  \citenamefont {Leefer}, \citenamefont {Jarmola}, \citenamefont {Dumeige},
  \citenamefont {M}, \citenamefont {Acosta}, \citenamefont {Kehayias},
  \citenamefont {Patton},\ and\ \citenamefont {Budker}}]{2014.PRL.Budker}%
  \BibitemOpen
  \bibfield  {author} {\bibinfo {author} {\bibfnamefont {K.}~\bibnamefont
  {Jensen}}, \bibinfo {author} {\bibfnamefont {N.}~\bibnamefont {Leefer}},
  \bibinfo {author} {\bibfnamefont {A.}~\bibnamefont {Jarmola}}, \bibinfo
  {author} {\bibfnamefont {Y.}~\bibnamefont {Dumeige}}, \bibinfo {author}
  {\bibfnamefont {V.}~\bibnamefont {M}}, \bibinfo {author} {\bibnamefont
  {Acosta}}, \bibinfo {author} {\bibfnamefont {P.}~\bibnamefont {Kehayias}},
  \bibinfo {author} {\bibfnamefont {B.}~\bibnamefont {Patton}}, \ and\ \bibinfo
  {author} {\bibfnamefont {D.}~\bibnamefont {Budker}},\ }\href@noop {}
  {\bibfield  {journal} {\bibinfo  {journal} {PRL}\ }\textbf {\bibinfo {volume}
  {112}},\ \bibinfo {pages} {160802} (\bibinfo {year} {2014})}\BibitemShut
  {NoStop}%
\bibitem [{\citenamefont {Ajoy}\ and\ \citenamefont
  {Cappellaro}(2012)}]{PhysRevA.86.062104}%
  \BibitemOpen
  \bibfield  {author} {\bibinfo {author} {\bibfnamefont {A.}~\bibnamefont
  {Ajoy}}\ and\ \bibinfo {author} {\bibfnamefont {P.}~\bibnamefont
  {Cappellaro}},\ }\href {\doibase 10.1103/PhysRevA.86.062104} {\bibfield
  {journal} {\bibinfo  {journal} {Phys. Rev. A}\ }\textbf {\bibinfo {volume}
  {86}},\ \bibinfo {pages} {062104} (\bibinfo {year} {2012})}\BibitemShut
  {NoStop}%
\bibitem [{\citenamefont {Hodges}\ \emph {et~al.}(2013)\citenamefont {Hodges},
  \citenamefont {Yao}, \citenamefont {Maclaurin}, \citenamefont {Rastogi},
  \citenamefont {Lukin},\ and\ \citenamefont {Englund}}]{PhysRevA.87.032118}%
  \BibitemOpen
  \bibfield  {author} {\bibinfo {author} {\bibfnamefont {J.~S.}\ \bibnamefont
  {Hodges}}, \bibinfo {author} {\bibfnamefont {N.~Y.}\ \bibnamefont {Yao}},
  \bibinfo {author} {\bibfnamefont {D.}~\bibnamefont {Maclaurin}}, \bibinfo
  {author} {\bibfnamefont {C.}~\bibnamefont {Rastogi}}, \bibinfo {author}
  {\bibfnamefont {M.~D.}\ \bibnamefont {Lukin}}, \ and\ \bibinfo {author}
  {\bibfnamefont {D.}~\bibnamefont {Englund}},\ }\href {\doibase
  10.1103/PhysRevA.87.032118} {\bibfield  {journal} {\bibinfo  {journal} {Phys.
  Rev. A}\ }\textbf {\bibinfo {volume} {87}},\ \bibinfo {pages} {032118}
  (\bibinfo {year} {2013})}\BibitemShut {NoStop}%
\bibitem [{\citenamefont {Pham}\ \emph {et~al.}(2011)\citenamefont {Pham},
  \citenamefont {Sage}, \citenamefont {Stanwix}, \citenamefont {Yeung},
  \citenamefont {Glenn}, \citenamefont {Trifonov}, \citenamefont {Cappellaro},
  \citenamefont {Hemmer}, \citenamefont {Lukin}, \citenamefont {Park},
  \citenamefont {Yacoby},\ and\ \citenamefont
  {Walsworth}}]{1367-2630-13-4-045021}%
  \BibitemOpen
  \bibfield  {author} {\bibinfo {author} {\bibfnamefont {L.~M.}\ \bibnamefont
  {Pham}}, \bibinfo {author} {\bibfnamefont {D.~L.}\ \bibnamefont {Sage}},
  \bibinfo {author} {\bibfnamefont {P.~L.}\ \bibnamefont {Stanwix}}, \bibinfo
  {author} {\bibfnamefont {T.~K.}\ \bibnamefont {Yeung}}, \bibinfo {author}
  {\bibfnamefont {D.}~\bibnamefont {Glenn}}, \bibinfo {author} {\bibfnamefont
  {A.}~\bibnamefont {Trifonov}}, \bibinfo {author} {\bibfnamefont
  {P.}~\bibnamefont {Cappellaro}}, \bibinfo {author} {\bibfnamefont {P.~R.}\
  \bibnamefont {Hemmer}}, \bibinfo {author} {\bibfnamefont {M.~D.}\
  \bibnamefont {Lukin}}, \bibinfo {author} {\bibfnamefont {H.}~\bibnamefont
  {Park}}, \bibinfo {author} {\bibfnamefont {A.}~\bibnamefont {Yacoby}}, \ and\
  \bibinfo {author} {\bibfnamefont {R.~L.}\ \bibnamefont {Walsworth}},\ }\href
  {http://stacks.iop.org/1367-2630/13/i=4/a=045021} {\bibfield  {journal}
  {\bibinfo  {journal} {New Journal of Physics}\ }\textbf {\bibinfo {volume}
  {13}},\ \bibinfo {pages} {045021} (\bibinfo {year} {2011})}\BibitemShut
  {NoStop}%
\bibitem [{\citenamefont {Manson}\ \emph {et~al.}(2006)\citenamefont {Manson},
  \citenamefont {Harrison},\ and\ \citenamefont
  {Sellars}}]{PhysRevB.74.104303}%
  \BibitemOpen
  \bibfield  {author} {\bibinfo {author} {\bibfnamefont {N.~B.}\ \bibnamefont
  {Manson}}, \bibinfo {author} {\bibfnamefont {J.~P.}\ \bibnamefont
  {Harrison}}, \ and\ \bibinfo {author} {\bibfnamefont {M.~J.}\ \bibnamefont
  {Sellars}},\ }\href {\doibase 10.1103/PhysRevB.74.104303} {\bibfield
  {journal} {\bibinfo  {journal} {Phys. Rev. B}\ }\textbf {\bibinfo {volume}
  {74}},\ \bibinfo {pages} {104303} (\bibinfo {year} {2006})}\BibitemShut
  {NoStop}%
\bibitem [{\citenamefont {Lin}\ \emph {et~al.}(2008)\citenamefont {Lin},
  \citenamefont {Wang}, \citenamefont {Chang}, \citenamefont {Hayashi},\ and\
  \citenamefont {Lin}}]{:/content/aip/journal/jcp/129/12/10.1063/1.2987717}%
  \BibitemOpen
  \bibfield  {author} {\bibinfo {author} {\bibfnamefont {C.-K.}\ \bibnamefont
  {Lin}}, \bibinfo {author} {\bibfnamefont {Y.-H.}\ \bibnamefont {Wang}},
  \bibinfo {author} {\bibfnamefont {H.-C.}\ \bibnamefont {Chang}}, \bibinfo
  {author} {\bibfnamefont {M.}~\bibnamefont {Hayashi}}, \ and\ \bibinfo
  {author} {\bibfnamefont {S.~H.}\ \bibnamefont {Lin}},\ }\href {\doibase
  http://dx.doi.org/10.1063/1.2987717} {\bibfield  {journal} {\bibinfo
  {journal} {The Journal of Chemical Physics}\ }\textbf {\bibinfo {volume}
  {129}},\ \bibinfo {eid} {124714} (\bibinfo {year} {2008})}\BibitemShut
  {NoStop}%
\bibitem [{\citenamefont {Wang}\ and\ \citenamefont
  {Takahashi}(2013)}]{PhysRevB.87.115122}%
  \BibitemOpen
  \bibfield  {author} {\bibinfo {author} {\bibfnamefont {Z.-H.}\ \bibnamefont
  {Wang}}\ and\ \bibinfo {author} {\bibfnamefont {S.}~\bibnamefont
  {Takahashi}},\ }\href {\doibase 10.1103/PhysRevB.87.115122} {\bibfield
  {journal} {\bibinfo  {journal} {Phys. Rev. B}\ }\textbf {\bibinfo {volume}
  {87}},\ \bibinfo {pages} {115122} (\bibinfo {year} {2013})}\BibitemShut
  {NoStop}%
\bibitem [{\citenamefont {Bar-Gill}\ \emph {et~al.}(2012)\citenamefont
  {Bar-Gill}, \citenamefont {Pham}, \citenamefont {Belthangady}, \citenamefont
  {Le~Sage}, \citenamefont {Cappellaro}, \citenamefont {Maze}, \citenamefont
  {Lukin}, \citenamefont {Yacoby},\ and\ \citenamefont
  {Walsworth}}]{Bar-Gill:2012fk}%
  \BibitemOpen
  \bibfield  {author} {\bibinfo {author} {\bibfnamefont {N.}~\bibnamefont
  {Bar-Gill}}, \bibinfo {author} {\bibfnamefont {L.~M.}\ \bibnamefont {Pham}},
  \bibinfo {author} {\bibfnamefont {C.}~\bibnamefont {Belthangady}}, \bibinfo
  {author} {\bibfnamefont {D.}~\bibnamefont {Le~Sage}}, \bibinfo {author}
  {\bibfnamefont {P.}~\bibnamefont {Cappellaro}}, \bibinfo {author}
  {\bibfnamefont {J.~R.}\ \bibnamefont {Maze}}, \bibinfo {author}
  {\bibfnamefont {M.~D.}\ \bibnamefont {Lukin}}, \bibinfo {author}
  {\bibfnamefont {A.}~\bibnamefont {Yacoby}}, \ and\ \bibinfo {author}
  {\bibfnamefont {R.}~\bibnamefont {Walsworth}},\ }\href
  {http://dx.doi.org/10.1038/ncomms1856} {\bibfield  {journal} {\bibinfo
  {journal} {Nat Commun}\ }\textbf {\bibinfo {volume} {3}} (\bibinfo {year}
  {2012})}\BibitemShut {NoStop}%
\bibitem [{\citenamefont {Narimanov}\ \emph {et~al.}(2005)\citenamefont
  {Narimanov}, \citenamefont {Fan},\ and\ \citenamefont
  {Gmachl}}]{narimanov2005compact}%
  \BibitemOpen
  \bibfield  {author} {\bibinfo {author} {\bibfnamefont {E.}~\bibnamefont
  {Narimanov}}, \bibinfo {author} {\bibfnamefont {J.}~\bibnamefont {Fan}}, \
  and\ \bibinfo {author} {\bibfnamefont {C.}~\bibnamefont {Gmachl}},\ }in\
  \href@noop {} {\emph {\bibinfo {booktitle} {Quantum Electronics and Laser
  Science Conference}}}\ (\bibinfo {organization} {Optical Society of
  America},\ \bibinfo {year} {2005})\ p.\ \bibinfo {pages} {QWA7}\BibitemShut
  {NoStop}%
\bibitem [{\citenamefont {Wrachtrup}\ and\ \citenamefont
  {Jelezko}(2006)}]{0953-8984-18-21-S08}%
  \BibitemOpen
  \bibfield  {author} {\bibinfo {author} {\bibfnamefont {J.}~\bibnamefont
  {Wrachtrup}}\ and\ \bibinfo {author} {\bibfnamefont {F.}~\bibnamefont
  {Jelezko}},\ }\href {http://stacks.iop.org/0953-8984/18/i=21/a=S08}
  {\bibfield  {journal} {\bibinfo  {journal} {Journal of Physics: Condensed
  Matter}\ }\textbf {\bibinfo {volume} {18}},\ \bibinfo {pages} {S807}
  (\bibinfo {year} {2006})}\BibitemShut {NoStop}%
\bibitem [{\citenamefont {Schr{\"o}der}\ \emph {et~al.}(2011)\citenamefont
  {Schr{\"o}der}, \citenamefont {G{\"a}deke}, \citenamefont {Banholzer},\ and\
  \citenamefont {Benson}}]{schroder2011ultrabright}%
  \BibitemOpen
  \bibfield  {author} {\bibinfo {author} {\bibfnamefont {T.}~\bibnamefont
  {Schr{\"o}der}}, \bibinfo {author} {\bibfnamefont {F.}~\bibnamefont
  {G{\"a}deke}}, \bibinfo {author} {\bibfnamefont {M.~J.}\ \bibnamefont
  {Banholzer}}, \ and\ \bibinfo {author} {\bibfnamefont {O.}~\bibnamefont
  {Benson}},\ }\href@noop {} {\bibfield  {journal} {\bibinfo  {journal} {New
  Journal of Physics}\ }\textbf {\bibinfo {volume} {13}},\ \bibinfo {pages}
  {055017} (\bibinfo {year} {2011})}\BibitemShut {NoStop}%
\bibitem [{\citenamefont {Acosta}\ \emph
  {et~al.}(2010{\natexlab{b}})\citenamefont {Acosta}, \citenamefont {Bauch},
  \citenamefont {Ledbetter}, \citenamefont {Waxman}, \citenamefont {Bouchard},\
  and\ \citenamefont {Budker}}]{2010.PRL.Budker.temperature_dependence}%
  \BibitemOpen
  \bibfield  {author} {\bibinfo {author} {\bibfnamefont {V.~M.}\ \bibnamefont
  {Acosta}}, \bibinfo {author} {\bibfnamefont {E.}~\bibnamefont {Bauch}},
  \bibinfo {author} {\bibfnamefont {M.~P.}\ \bibnamefont {Ledbetter}}, \bibinfo
  {author} {\bibfnamefont {A.}~\bibnamefont {Waxman}}, \bibinfo {author}
  {\bibfnamefont {L.-S.}\ \bibnamefont {Bouchard}}, \ and\ \bibinfo {author}
  {\bibfnamefont {D.}~\bibnamefont {Budker}},\ }\href {\doibase
  10.1103/PhysRevLett.104.070801} {\bibfield  {journal} {\bibinfo  {journal}
  {Phys. Rev. Lett.}\ }\textbf {\bibinfo {volume} {104}},\ \bibinfo {pages}
  {070801} (\bibinfo {year} {2010}{\natexlab{b}})}\BibitemShut {NoStop}%
\bibitem [{\citenamefont {Fang}\ \emph
  {et~al.}(2013{\natexlab{b}})\citenamefont {Fang}, \citenamefont {Acosta},
  \citenamefont {Santori}, \citenamefont {Huang}, \citenamefont {Itoh},
  \citenamefont {Watanabe}, \citenamefont {Shikata},\ and\ \citenamefont
  {Beausoleil}}]{2013.Beausoleil.PRL}%
  \BibitemOpen
  \bibfield  {author} {\bibinfo {author} {\bibfnamefont {K.}~\bibnamefont
  {Fang}}, \bibinfo {author} {\bibfnamefont {V.~M.}\ \bibnamefont {Acosta}},
  \bibinfo {author} {\bibfnamefont {C.}~\bibnamefont {Santori}}, \bibinfo
  {author} {\bibfnamefont {Z.}~\bibnamefont {Huang}}, \bibinfo {author}
  {\bibfnamefont {K.~M.}\ \bibnamefont {Itoh}}, \bibinfo {author}
  {\bibfnamefont {H.}~\bibnamefont {Watanabe}}, \bibinfo {author}
  {\bibfnamefont {S.}~\bibnamefont {Shikata}}, \ and\ \bibinfo {author}
  {\bibfnamefont {R.~G.}\ \bibnamefont {Beausoleil}},\ }\href {\doibase
  10.1103/PhysRevLett.110.130802} {\bibfield  {journal} {\bibinfo  {journal}
  {Phys. Rev. Lett.}\ }\textbf {\bibinfo {volume} {110}},\ \bibinfo {pages}
  {130802} (\bibinfo {year} {2013}{\natexlab{b}})}\BibitemShut {NoStop}%
\bibitem [{\citenamefont {Matushita}\ and\ \citenamefont
  {Campbell}(1967)}]{1967geomagnetics}%
  \BibitemOpen
  \bibinfo {editor} {\bibfnamefont {S.}~\bibnamefont {Matushita}}\ and\
  \bibinfo {editor} {\bibfnamefont {W.~H.}\ \bibnamefont {Campbell}},\ eds.,\
  \href@noop {} {\emph {\bibinfo {title} {Physics of Geomagnetic Phenomena}}}\
  (\bibinfo  {publisher} {Acacemic Press},\ \bibinfo {year} {1967})\BibitemShut
  {NoStop}%
\end{thebibliography}

\section*{Acknowledgements}
The authors would like to thank R.\,J. Shiue, D. Rich, G. Steinbrecher, E.\,H. Chen, and O. Gaathon for helpful discussion. The authors would like to acknowledge D. Twitchen and M. Markham at Element Six Ltd. for assistance with diamond irradiation. This work was supported by The Defense Advanced Research Projects Agency (government contract/grant number N66001-13-1-4027). The views, opinions, and/or findings contained in this article are those of the authors and should not be interpreted as representing the official views or policies, either expressed or implied, of the Defense Advanced Research Projects Agency or the Department of Defense. The Lincoln Laboratory portion of this work is sponsored by the Assistant Secretary of Defense for Research \& Engineering under Air Force Contract \#FA8721-05-C-0002 and the Office of Naval Research Section 219. Opinions, interpretations, conclusions and recommendations are those of the authors and are not necessarily endorsed by the United States Government. M.\,T. was supported by the NSF IGERT program Interdisciplinary Quantum Information Science and Engineering (iQuiSE). T.S. was supported by the Alexander von Humboldt Foundation. H.\,C. is supported by the NASA Office of the Chief Technologist's Space Technology Research Fellowship. 

\section*{Author Contributions}
H.\,C., M.\,E.\,T., T.\,S., D.\,B. and D.\,E. conceived and designed the experiments. H.\,C., C.\,T., and D.\,B. performed the experiments. H.\,C., D.\,B., and D.\,E. prepared the manuscript. All authors reviewed the manuscript. 

\section*{Competing Financial Interests Statement}
The authors declare no competing financial interests.

\end{document}